\documentclass{article}

\PassOptionsToPackage{numbers, compress}{natbib}
\usepackage[final]{nips_2017}

\usepackage[utf8]{inputenc} 
\usepackage[T1]{fontenc}    
\usepackage{hyperref}       
\usepackage{url}            
\usepackage{booktabs}       
\usepackage{amsfonts}       
\usepackage{nicefrac}       
\usepackage{microtype}      
\usepackage{graphicx}

\title{Evaluating deep variational autoencoders trained on pan-cancer gene expression}

\author{
  Gregory P. Way \\
  Genomics and Computational Biology Graduate Group \\
  University of Pennsylvania \\
  Philadelphia, PA 19143 \\
  \texttt{gregory.way@gmail.com} \\
  \AND
  Casey S. Greene \thanks{Corresponding Author} \\
  Department of Systems Pharmacology and Translational Therapeutics \\
  University of Pennsylvania \\
  Philadelphia, PA 19143 \\
  \texttt{greenescientist@gmail.com}
}

\begin{document}

\maketitle

\begin{abstract} Cancer is a heterogeneous disease with diverse molecular
etiologies and outcomes. The Cancer Genome Atlas (TCGA) has released a large
compendium of over 10,000 tumors with RNA-seq gene expression measurements. Gene
expression captures the diverse molecular profiles of tumors and can be
interrogated to reveal differential pathway activations. Deep unsupervised
models, including Variational Autoencoders (VAE) can be used to reveal these
underlying patterns. We compare a one-hidden layer VAE to two alternative VAE
architectures with increased depth. We determine the additional capacity
marginally improves performance. We train and compare the three VAE
architectures to other dimensionality reduction techniques including principal
components analysis (PCA), independent components analysis (ICA), non-negative
matrix factorization (NMF), and analysis of gene expression by denoising
autoencoders (ADAGE). We compare performance in a supervised learning task
predicting gene inactivation pan-cancer and in a latent space analysis of high
grade serous ovarian cancer (HGSC) subtypes. We do not observe substantial
differences across algorithms in the classification task. VAE latent spaces
offer biological insights into HGSC subtype biology. \end{abstract}

\section{Introduction}

From a systems biology perspective, the transcriptome can reveal the
overall state of a tumor \cite{huang_cancer_2009}. This state involves diverse
etiologies and aberrantly active pathways that act together to produce
neoplasia, growth, and metastasis \cite{hanahan_hallmarks_2011}. Such patterns
can be extracted with machine learning.

Deep generative models have improved state of the art in several domains
including image and text processing
\cite{kingma_auto-encoding_2013,rezende_stochastic_2014,goodfellow_generative_2014}.
Such models can simulate realistic data by learning an underlying data
generating manifold. The manifold can be mathematically manipulated to extract
interpretable elements from the data. For example, a recent imaging study
subtracted the vector representation of a ``neutral woman'' from a ``smiling
woman'', added the result to a ``neutral man'', and revealed image vectors
representing ``smiling men'' \cite{radford_unsupervised_2015}. In text
processing, $\vec{king} - \vec{man} + \vec{woman} = \vec{queen}$
\cite{mikolov_efficient_2013}.

Previously, nonlinear dimensionality reduction approaches have revealed complex
patterns and novel biology from publicly available gene expression data
\cite{chen_learning_2016,tan_unsupervised_2017}, including drug response
predictions \cite{rampasek_dr.vae:_2017}. Here, we extend a one-hidden layer VAE
model, named Tybalt \cite{way_extracting_2017}. We train and evaluate Tybalt,
plus two other VAE architectures and compare them to other dimensionality
reduction algorithms.

\section{Methods}

\subsection{The Cancer Genome Atlas RNAseq Data}

We used TCGA PanCanAtlas RNA-seq data from 33 different cancer-types
\cite{weinstein_cancer_2013}. The data includes 10,459 samples (9,732 tumors and
727 tumor adjacent normal). The data was batch corrected and RSEM preprocessed,
and is in the log2(FPKM + 1) format. To facilitate VAE training, we zero-one
normalized by gene. We also used zero-one normalization for NMF and ADAGE. We
used z-scored data for PCA and ICA. All data are publicly available and were
accessed from the UCSC Xena browser under a versioned Zenodo archive
\cite{way_data_2016}.

\subsection{Variational Autoencoder Training}

We trained VAE models as previously described \cite{way_extracting_2017}. We
performed a grid search over hyperparameters and defined optimal models by
lowest holdout validation loss. VAE loss is the sum of reconstruction loss and a
Kullback-Leibler (KL) divergence term constraining feature activations to a
Gaussian distribution \cite{kingma_auto-encoding_2013,rezende_stochastic_2014}.
We searched over batch sizes, epochs, learning rates, and kappa values. Kappa
controls ``warmup'', which determines how quickly the loss term incorporates KL
divergence \cite{sonderby_ladder_2016}. VAEs learn two distinct latent
representations, a mean and a standard deviation vector, which are
reparametrized into a single vector that can be back-propagated. This enables
rapid sampling over features to simulate data.We trained our models using Keras
\cite{chollet_keras_2015} with a TensorFlow backend
\cite{abadi_tensorflow:_2016}. We trained and evaluated the performance of three
VAE architectures (Table \ref{gpw:table1}). Using an EVGA GeForce GTX 1060 GPU,
all VAE models were trained in under 4 minutes.

\begin{table}[h]
  \caption{VAE Architectures}
  \label{gpw:table1}
  \centering
  \begin{tabular}{llll}
    \toprule
    Name & Hidden Layers & Hidden Layer Size & Latent Feature Size \\
    \midrule
    Tybalt & 1 &  & 100     \\
    Two Hidden VAE (100) & 2 & 100 & 100      \\
    Two Hidden VAE (300) & 2 & 300 & 100 \\
    \bottomrule
  \end{tabular}
\end{table}

\subsection{Dimensionality Reduction Analysis}

We evaluated the ability of the VAE models to generate biologically meaningful
features. We compared these features to PCA, ICA, NMF, and ADAGE features
derived from the same pan-cancer RNAseq data.

\subsubsection{Predicting NF1 inactivation in tumors from gene expression data}

Molecular aberrations in cancer form gene expression signatures that supervised
machine learning algorithms can detect
\cite{guinney_modeling_2014,way_machine_2017}. We trained elastic net logistic
regression classifiers to detect pan-cancer tumors with inactivated NF1. We
defined NF1 inactivated tumors based on the presence of deleterious NF1
mutations or NF1 deep copy number loss. NF1 inactivation is difficult but
important to detect because NF1 can be inactivated by multiple mechanisms and
reliable classification is required for effective targeted therapies
\cite{mcgillicuddy_proteasomal_2009}. We trained independent models using
pan-cancer RNAseq features derived from each dimensionality reduction algorithm.
We report performance during 5-fold cross validation. We performed all training
and evaluation using sci-kit learn \cite{pedregosa_scikit-learn:_2011}. We
trained our models using tumors that had matched mutation, copy number, and
RNAseq data, and we used only cancer-types that had greater than 10 NF1
inactivated samples ($n = 1774$). This included bladder carcinoma (BLCA), low
grade glioma (LGG), lung adenocarcinoma (LUAD), paraganglioma and
pheochromocytoma (PCPG), skin cutaneous melanoma (SKCM), and stomach
adenocarcinoma (STAD). As a negative control, we shuffled gene expression
profiles for each sample independently, and used these shuffled matrices for
predictions.

\subsubsection{High grade serous ovarian cancer arithmetic}

Four HGSC subtypes have been previously described
\cite{network_integrated_2011}. However, they are not consistent across
populations \cite{way_comprehensive_2016}. Using original TCGA subtype labels
\cite{verhaak_prognostically_2012}, we calculated mean HGSC subtype vector
representations for the mesenchymal and immunoreactive subtypes with the latent
space representation from each dimensionality reduction algorithm. Samples from
the two subtypes were often aggregated under different clustering
initializations \cite{way_comprehensive_2016}. The mesenchymal subtype is
defined by poor prognoses, overexpression of extracellular matrix genes, and
increased desmoplasia, while the immunoreactive subtype displays increased
survival and immune cell infiltration
\cite{tothill_novel_2008,konecny_prognostic_2014}. We hypothesized that
subtracting latent space representations of subtypes would reveal biological
patterns. To characterize the patterns, we performed pathway overrepresentation
analyses (ORA) \cite{wang_webgestalt_2017} over Gene Ontology (GO) terms
\cite{ashburner_gene_2000} using the high weight genes ($> 2.5$ standard
deviations from the mean) of the highest differentiating positive feature. We
report the most significantly overrepresented GO term.

\section{Results}

\subsection{Modest improvements by architecture}

\begin{figure}[!ht] \centerline{\includegraphics[width=2.4in]{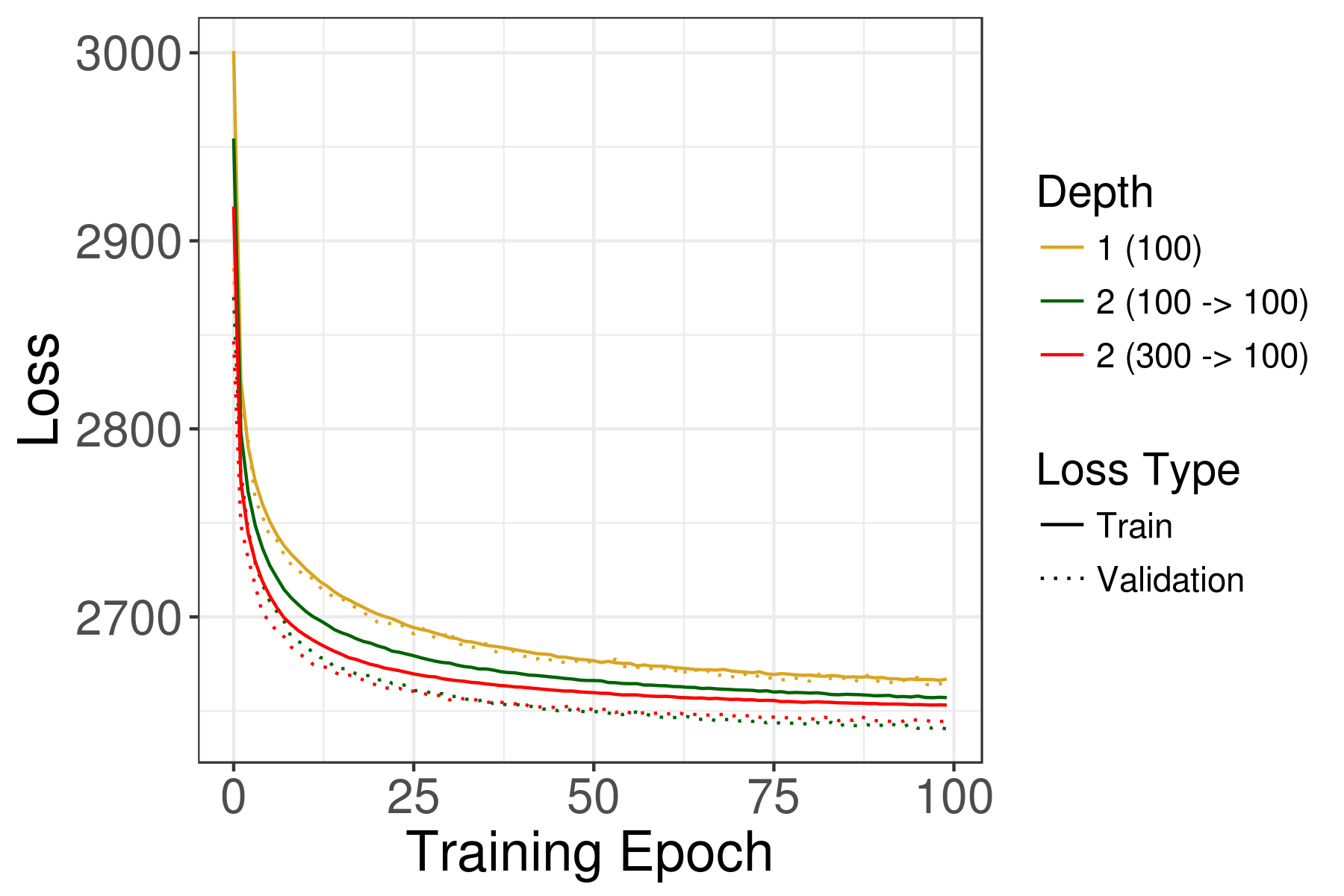}}
\caption{Evaluating three VAE architectures reveals modest performance
improvements for deeper models. Models were selected based on lowest holdout
validation loss at training end. Training was relatively stable for many
hyperparameter combinations. Random fluctuations and selection bias cause
validation loss to appear, counterintuitively, lower than training
loss.}\label{gpw:fig1} \end{figure}

We observed increased performance for the two-hidden layer models, and an
additional increase for the model with two compression steps. However, the
benefits were modest as compared to the one-hidden layer Tybalt model (Figure
\ref{gpw:fig1}).

\subsection{Dimensionality Reduction Comparison}

\subsubsection{Supervised classification of NF1 inactivation}

\begin{figure}[!ht] \centerline{\includegraphics[width=3.1in]{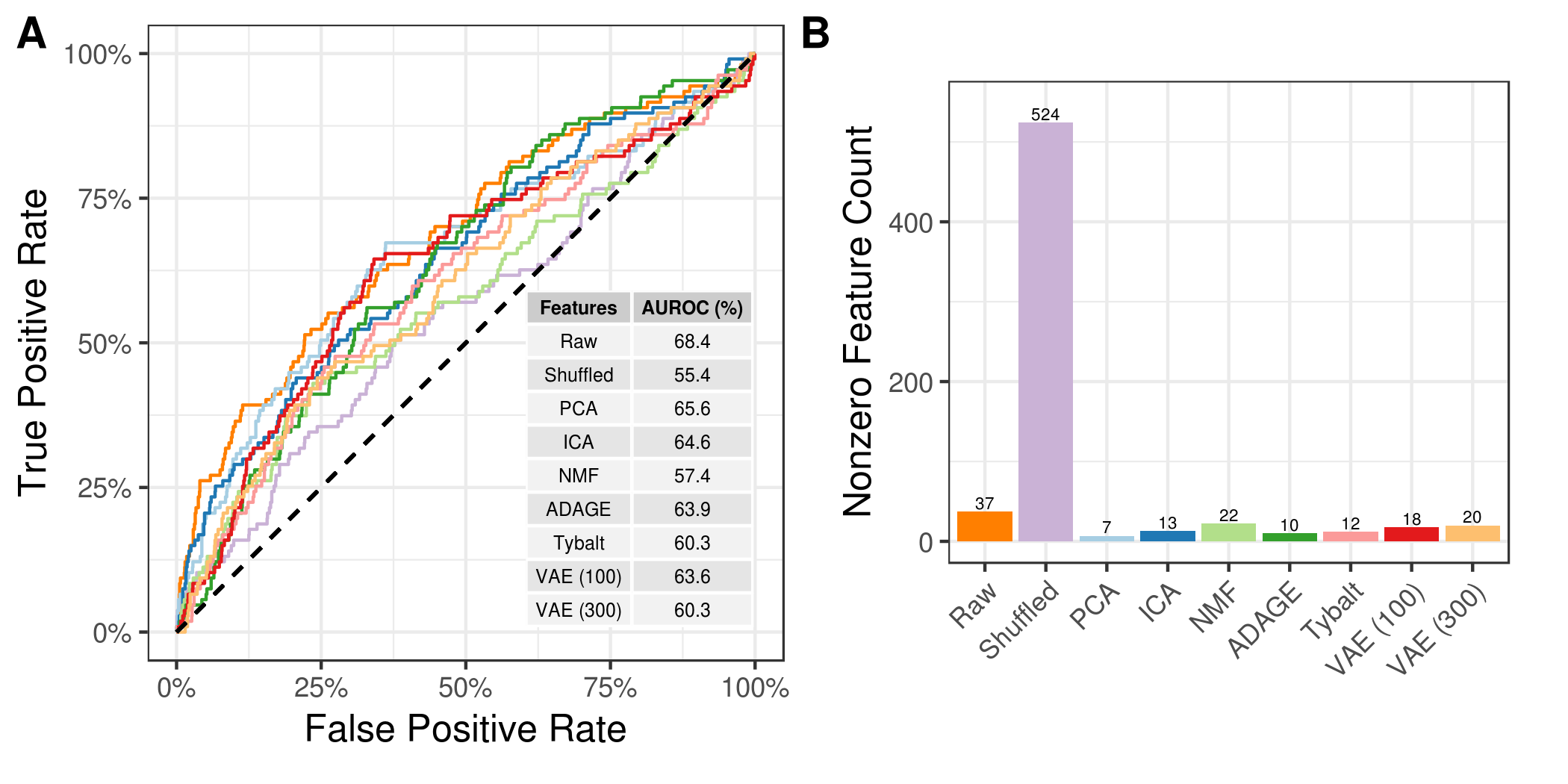}}
\caption{Evaluating dimensionality reduction algorithms in predicting NF1
inactivation pan-cancer. \textbf{(A)} Receiver operating characteristic curve
for cross validation intervals. \textbf{(B)} Number of features selected by each
model. The colors are consistent between figure panels.}\label{gpw:fig2}
\end{figure}

Based on receiver operating characteristic (ROC) curves, all algorithms had
relatively similar performance (Figure \ref{gpw:fig2}A). PCA performed slightly
better than other dimensionality reduction algorithms (area under the ROC curve
(AUROC) = 65.6\%). However, raw gene expression features performed the best
(AUROC = 68.4\%). There were more classification features used for models built
with raw features as compared to other methods; NMF included many features where
PCA included the fewest (Figure \ref{gpw:fig2}B). The shuffled dataset contained
the highest number of features, and was somewhat predictive of NF1 status (AUROC =
55.4\%), which may be an artifact of relatively low sample sizes.

\subsubsection{Latent space arithmetic of HGSC subtypes}

\begin{figure}[!ht] \centerline{\includegraphics[width=2.3in]{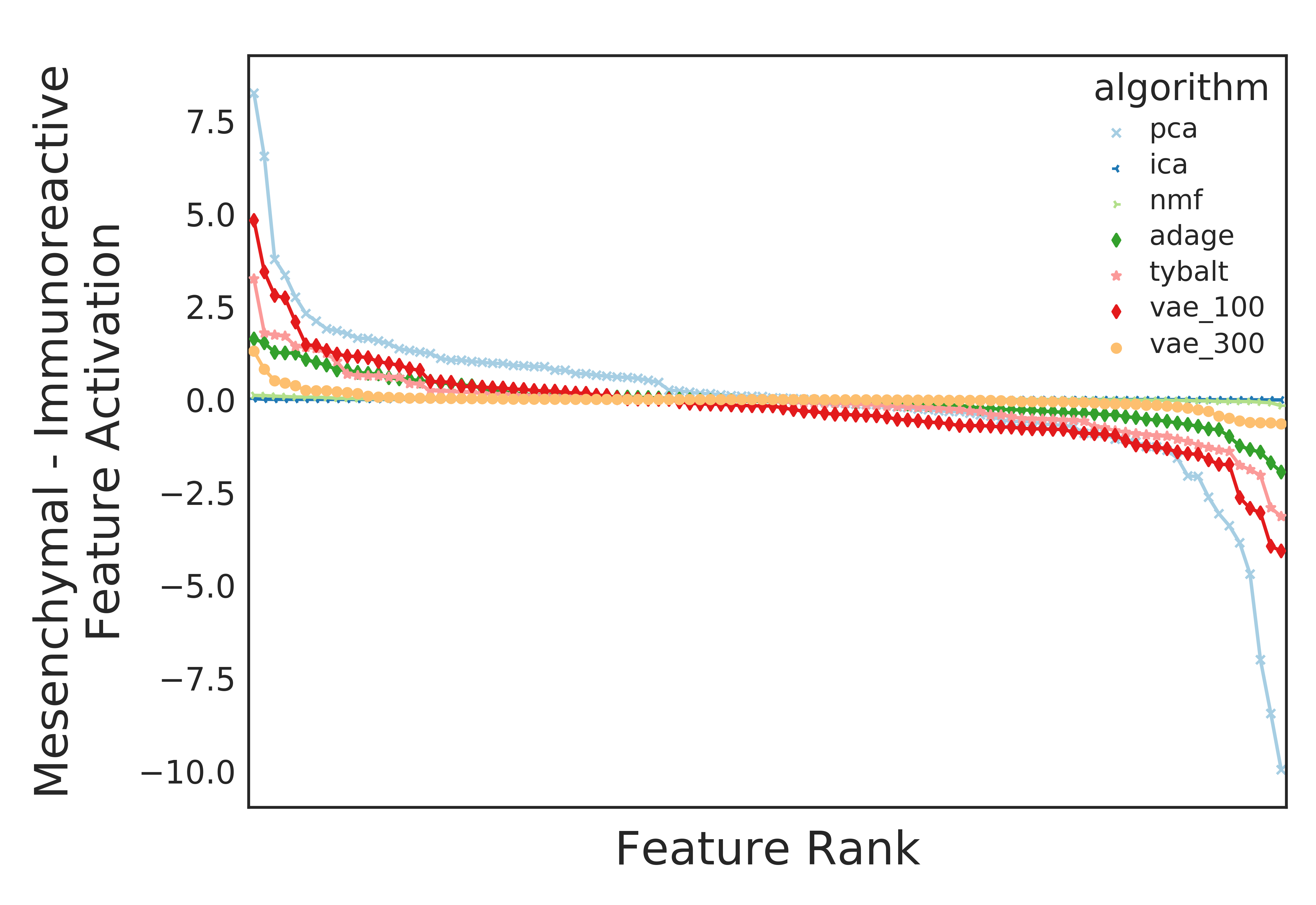}}
\caption{Subtracting mean vector representations of the Mesenchymal and
Immunoreactive High Grade Serous Ovarian Cancer (HGSC)
subtypes.}\label{gpw:fig3} \end{figure}

We ranked mean activation differences across dimensionality reduction methods
(Figure \ref{gpw:fig3}). PCA included the largest number of features with high
values, while ICA and NMF had few activation differences. The nonlinear neural
network based approaches displayed sparse feature differences. ORA analyses
revealed few significant terms in the linear methods (Table \ref{gpw:table2}).
Of the nonlinear methods, Tybalt, and the 300-hidden node VAE both identified a
collagen term, which is known to be associated with mesenchymal subtype tumors
\cite{tothill_novel_2008}.

\begin{table}[!ht]
  \caption{Dimensionality reduction algorithms subtraction comparison}
\label{gpw:table2}
  \centering
  \begin{tabular}{lll}
    \toprule
    Algorithm & Top Pathway & Adj. p value \\
    \midrule
    PCA & \textit{Zero high weight genes} & \\
    ICA & \textit{No significant pathways} & \\
    NMF & Homophilic Cell Adhesion Via Plasma Membrane Adhesion Molecules & $3.3e^{-06}$ \\
    ADAGE & \textit{No significant pathways} & \\
    Tybalt & Collagen Catabolic Process & $1.8e^{-09}$ \\
    VAE (100) & Epidermis Development & $8.0e^{-04}$ \\
    VAE (300) & Collagen Catabolic Process & $1.7e^{-03}$ \\
    \bottomrule
  \end{tabular}
\end{table}

\section{Conclusions}

We evaluated the performance of three VAE models trained on TCGA pan-cancer gene
expression. While we did not explore larger architectures with higher capacity,
it appears that increasing the depth of model only modestly improves
performance. It is also likely that increasing model depth reduces the ability
to interpret the model. We also did not compare performance across different
sized latent spaces. Nevertheless, we show that VAEs capture signals that are
able to predict gene inactivation comparably to other algorithms. We demonstrate
that the VAE latent space arithmetic provides a unique benefit and should be
explored further in the context of gene expression data. We provide source code
to reproduce training and latent space analyses at
https://github.com/greenelab/tybalt \cite{way_greenelab/tybalt:_2017} and
pan-cancer classifier analyses at https://github.com/greenelab/pancancer
\cite{way_greenelab/pancancer:_2017}.

\subsubsection*{Acknowledgments}

This work was supported by NIH grant T32 HG000046 (GPW) and GBMF 4552 from the
Gordon and Betty Moore Foundation (CSG). We would like to thank Jaclyn N.
Taroni, David Nicholson, and Daniel Himmelstein for code review.

\medskip

\small

\bibliographystyle{unsrtnat}
\bibliography{nips}

\end{document}